\documentclass{ws-procs9x6}

\def\today{4~January~2010}

\def \jmax{J}
\def \bfv {{\bf v}}
\def \cv  {c_{\rm v}}
\def \cz  {c_{\rm z}}
\def \cnl {c_{\rm NL}}
\def \vs{v_{\rm s}}
\def \uzero{u_{0,\varepsilon}} 
\def \vzero{v_{0,\varepsilon}} 
\def \Adim{\lambda} 
\def \tildeK  {\tilde{K}} 
\def \Alpha {A}
\def \Beta  {B}

\def \mod#1{\vert #1 \vert}
\def \div{\mathop{\rm div}\nolimits}
\def \grad{\mathop{\rm grad}\nolimits}
\def \sech{\mathop{\rm sech}\nolimits}
\def \coth{\mathop{\rm coth}\nolimits}

\begin{document}

\title{
EXACT INTERNAL WAVES OF A BOUSSINESQ SYSTEM\footnote{
\textit{Waves and stability in continuous media},
eds.~A.~Greco, S.~Rionero and T.~Ruggeri
(World scientific, Singapore, 2010).
WASCOM 15, Mondello (Pa), 28 June--1 July 2009.}}

\author{Hai Yen NGUYEN}
\address{
Laboratoire de physique des oc\'eans,
IFREMER, BP 70, F--29280 Plouzan\'e, France
\\ E-mail: mathsyen@yahoo.com
}
{\vglue -4.0 truemm}

\author{Fr\'ed\'eric DIAS}
\address{Centre de math\'ematiques et de leurs applications (UMR 8536),
\\ \'Ecole normale sup\'erieure de Cachan,
\\ 61, avenue du Pr\'esident Wilson, F--94235 Cachan Cedex, France.
\\ University College Dublin, School of Mathematical Sciences,
\\ Belfield, Dublin 4, Ireland
\\ E-mail: Frederic.Dias@cmla.ens-cachan.fr
}
{\vglue -4.0 truemm}

\author{Robert CONTE}
\address{LRC MESO (ENS Cachan et CEA-DAM),
\\ Centre de math\'ematiques et de leurs applications (UMR 8536),
\\ \'Ecole normale sup\'erieure de Cachan, 
\\ 61, avenue du Pr\'esident Wilson, F--94235 Cachan Cedex, France.
\\  
Service de physique de l'\'etat condens\'e (URA 2464)
\\ CEA--Saclay, F--91191 Gif-sur-Yvette Cedex, France
\\ E-mail:  Robert.Conte@cea.fr
}
{\vglue -4.0 truemm}


\begin{abstract}
We consider a Boussinesq system describing one-dimensional internal waves
which develop at the boundary between two immiscible fluids,
and we restrict to its traveling waves.
The method which yields explicitly all
the elliptic or degenerate elliptic 
solutions of a given nonlinear, any order algebraic ordinary differential equation
is briefly recalled.
We then apply it to the fluid system and,
restricting in this preliminary report to the generic situation,
we obtain all the solutions in that class,
including several new solutions. 
\end{abstract}


\noindent



\noindent

\bodymatter

\section{Introduction} 
\label{sectionIntro}

At the boundary between two immiscible fluids,
one observes the formation of waves, called \textit{internal waves}.
These are typically described by Boussinesq systems
such as \cite[Eq.~(28)]{BLS2008}
\begin{eqnarray}
& & {\hskip -10.0 truemm}
\left\lbrace
\begin{array}{ll}
\displaystyle{
(1- \mu b \Delta)\partial_t \zeta 
+ \cv \div \bfv + \cnl \div (\zeta \bfv) + \mu a \div \Delta \bfv=0,
}
\\
\displaystyle{ 
(1- \mu d \Delta)\partial_t \bfv + \cz \grad \zeta 
+ \frac{\cnl}{2} \grad\mod{\bfv}^2 + \mu c \Delta \grad \zeta=0,
}
\end{array}
\right.
\label{eqBLS28}
\end{eqnarray}
in which $\mu,a,b,c,d,\cnl,\cv,\cz$ are constant, $\mu \cnl\not=0$.

We restrict here to one-dimensional situations,
relevant for instance when the fluids are inside a channel,
and our purpose is to obtain traveling waves 
$\zeta=u(\xi)-\cv / \cnl,\bfv=v(\xi)+ c_0 / \cnl, \xi=x-c_0 t$ 
(the translation on $u$ suppresses any dependence on $\cv$
and the one on $v$ shortens the expressions below)
in closed form by a nonperturbative method.
The conservative form of the equations (\ref{eqBLS28}) allows
each equation to be integrated once,
and the considered system will be
\begin{eqnarray}
& & {\hskip -10.0 truemm}
\left\lbrace
\begin{array}{ll}
\displaystyle{
\mu b c_0 u'' + \mu a v'' + \cnl u v + K_1 =0,\
}
\\
\displaystyle{ 
\mu d c_0 v'' + \mu c u'' + \frac{\cnl}{2} v^2 + \cz u + K_2 =0.
}
\end{array}
\right.
\label{eqBLS28_xi}
\end{eqnarray}
The above system is essentially the same as those considered
by Chen \cite{Chen1998}
and Nguyen and Dias \cite[Eq.~(41)]{NguyenDias2008}.

Chen \cite{Chen1998} already found all the traveling waves 
in which $u$ and $v$ are polynomials
(of degree $1,2$ or $4$) in $\tanh k \xi$ and $\sech k \xi$.
In the present work, 
we obtain the closed form expressions
of \textit{all}\footnote{In this short report, only the generic case
is presented.}
those solutions of (\ref{eqBLS28_xi}) 
which are either elliptic
(doubly periodic in the $\xi$ complex plane)
or degenerate elliptic,
i.e.~rational in one exponential $e^{k \xi}$
(simply periodic in the $\xi$ complex plane,
which includes the above mentioned solutions)
or rational in $\xi$.

The method,
based on those complex singularities
of $(u,v)$ which depend on the initial conditions
(``movable'' singularities \cite{CMBook}), 
implements classical results by 
Briot, Bouquet \cite{BriotBouquet}
and Poincar\'e \cite{MapleAlgcurves}.
First presented in \cite{MC2003}, 
it was later turned into an algorithm \cite{CM2009}.

\section{Singularity analysis}
\label{sectionAnalysis}

In order to know whether closed-form solutions to (\ref{eqBLS28_xi})
might exist,
a prerequisite \cite{CMBook} is to investigate the singularities of $(u,v)$
in the complex plane of $\xi$.
One must distinguish whether the total differential order 
of system (\ref{eqBLS28_xi}) is four or two,
depending on the value of the determinant
\begin{eqnarray}
& & 
\delta \equiv b d c_0^2 - a c. 
\label{eqdefdelta}
\end{eqnarray}
We leave the nongeneric case $\delta=0$ to a forthcoming detailed study.

In the generic case $\delta\not=0$,
the system is equivalent to
\begin{eqnarray}
& & {\hskip -10.0 truemm}
\delta\not=0:\
\left\lbrace
\begin{array}{ll}
\displaystyle{
\mu \delta u''
 + \cnl \left(d c_0 u v - \frac{a}{2} v^2\right)
 - a \cz u + d c_0 K_1 - a K_2=0,
}
\\
\displaystyle{
\mu \delta v''
 + \cnl \left(\frac{b c_0}{2} v^2 - c u v\right)
 + b c_0 \cz u + b c_0 K_2 - c K_1=0.
}
\end{array}
\right.
\label{eqBLS28_xiGeneric}
\end{eqnarray}
For an easier computation of $u$ knowing $v$,
it is convenient to introduce the shift
\begin{eqnarray}
& & 
\vs = v - \frac{b c_0}{\cnl c} \cz.
\label{eqdefvs}
\end{eqnarray}
In order to find all the elliptic and 
degenerate elliptic solutions,
one must first determine all the families of movable poles
\textit{and} movable zeros (i.e.~movable poles of $1/u$ or $1/\vs$).

Let us first determine the poles. 
Assume that, near a movable singularity $\xi_0$, 
the variables $(u,v)$ behave algebraically
\begin{eqnarray}
& & {\hskip -10.0 truemm}
u \sim u_0 \chi^{p_1},\
v \sim v_0 \chi^{p_2},\
\chi=\xi-\xi_0,\ 
u_0 v_0 \not=0,
\label{eqLeading}
\end{eqnarray}
with $p_1,p_2$ not both positive integers.
Balancing the highest derivatives and the nonlinear terms,
one generically (nongeneric cases will be dealt with in a forthcoming paper) 
obtains double pole behaviours
\begin{eqnarray}
& & {\hskip -10.0 truemm}
\left\lbrace
\begin{array}{ll}
\displaystyle{
p_1=-2,\ p_2=-2,
}\\ \displaystyle{
      \cnl     u_0 v_0 +6 \mu (b c_0 u_0 + a v_0)=0,\
\frac{\cnl}{2} v_0^2   +6 \mu (d c_0 v_0 + c u_0)=0.
}
\end{array}
\right.
\label{eqLeadingGeneric}
\end{eqnarray}

Whenever $a c D \not=0$, with
\begin{eqnarray}
& & 
D^2 \equiv (b-2 d)^2 c_0^2+ 8 a c,\
\label{eqdefD}
\end{eqnarray}
this system (\ref{eqLeadingGeneric}) admits two solutions $(u_0,v_0)$,
\begin{eqnarray}
& & {\hskip -10.0 truemm}
\delta a c D \not=0:\
\left\lbrace
\begin{array}{ll}
\displaystyle{
\uzero=\frac{3 \mu}{2 c \cnl}
 \left(b (2 d -b) c_0^2 - 4 a c + \varepsilon b c_0 D\right),\
}\\ \displaystyle{ 
\vzero=\frac{3 \mu}{\cnl} \left(-(b + 2 d) c_0 + \varepsilon D\right),\ 
\varepsilon^2=1.
}
\end{array}
\right.
\label{eqLeadingGenericSol}
\end{eqnarray}
Let us next determine the movable zeros. 
By elimination between (\ref{eqBLS28_xiGeneric}),
it is easy to establish the fourth order ODE for $1/\vs$.
Its movable poles are one simple pole of arbitrary residue
plus, when $K_{11}\not=0$, one double pole, with
\begin{eqnarray}
& & 
K_{11} = K_1 - \frac{\Adim D - 2 d c_0}{c} 
           \left(K_2 + \frac{\cz^2}{2 c^2} (\Adim D -2 d c_0)^2\right).
 \end{eqnarray}
The movable zeros of $u$ are less easy to establish,
but it is sufficient for our purpose, as explained below,
to know that $u$ has always at least two movable zeros.
 
One must then compute \cite{CMBook}
the Fuchs indices $i$ of the linearized system of
(\ref{eqBLS28_xiGeneric}) near all the movable singularities. 
Near the movable double poles (\ref{eqLeading}), 
the resulting indicial equation (we skip the details)
only depends on one adimensional parameter $\Adim$,
\begin{eqnarray}
& & 
(i+1)(i-6) \left(i^2 - 5 i + \frac{12}{1 + \varepsilon \Adim}\right)=0,
\label{eqindices}
\\ & & 
\Adim=\frac{(b+2 d) c_0}{D}.
\label{eqdefAdim}
\end{eqnarray}
Since some Fuchs indices $i$ are generically noninteger,
the general solution of the system (\ref{eqBLS28_xiGeneric}) is multivalued.
Nongenerically, for the general solution to be singlevalued,
it is necessary that, for both signs $\varepsilon$,
all roots $i$ of (\ref{eqindices}) be integer.
Denoting these roots as
\begin{eqnarray}
& & {\hskip -10.0 truemm}
-1,6,\frac{5+Q}{2},\frac{5-Q}{2}\hbox{ for } \varepsilon=+1,\ 
-1,6,\frac{5+R}{2},\frac{5-R}{2}\hbox{ for } \varepsilon=-1,\
\end{eqnarray}
the elimination of $\Adim$ between the products of the roots 
\begin{eqnarray}
& & 
\frac{25-Q^2}{4}=\frac{12}{1+\Adim},\
\frac{25-R^2}{4}=\frac{12}{1-\Adim},\
\end{eqnarray}
yields the diophantine equation
\begin{eqnarray}
& & 
\frac{24}{25-Q^2} + \frac{24}{25-R^2} = 1,
\end{eqnarray}
which admits no solution for odd positive integers $(Q,R)$.

Despite its generically multivalued general solution,
the system (\ref{eqBLS28_xiGeneric}) may still 
admit singlevalued particular solutions.
For this it is necessary that the Laurent series whose first term is 
(\ref{eqLeading}),
\begin{eqnarray}
& & 
u=\sum_{j=0}^{+\infty} u_j \chi^{j-2},\
v=\sum_{j=0}^{+\infty} v_j \chi^{j-2},\
\end{eqnarray}
exists,
i.e.~that no impossibility occurs when computing the coefficients $u_j,v_j$.
The invariance of the system (\ref{eqBLS28_xi}) 
under $\xi \to -\xi$ forbids the occurrence of odd powers of $\chi$ 
in the Laurent series of $u$ and $v$.

For convenience, the three defining equations for $\delta,D, \Adim$ 
can be solved for $a, b c_0, \delta$ in terms of $d c_0, \Adim, D, c$,
yielding
\begin{eqnarray}
& & 
a=\frac{D^2 - (4 d c_0 - \Adim D)^2}{8 c},\
b c0=-2 d c_0 + \Adim D,\
\delta=(\Adim^2-1) D^2,
\\ & &
\uzero=\frac{3 \mu D}{4 c \cnl} (\Adim - \varepsilon) 
\left\lbrack 4 d c_0 - D (\Adim - \varepsilon)\right\rbrack,\
\vzero=\frac{3 \mu D}{\cnl} (\varepsilon - \Adim). 
\label{equ0v0}
\end{eqnarray}

For a generic $\Adim$,  the values $i=2$ and $i=4$ are not roots
of (\ref{eqindices}),
\begin{eqnarray}
& & 
\frac{(b+2d)^2 c_0^2}{(b-2d)^2 c_0^2 + 8 a c} \notin \lbrace 1,4\rbrace,
\label{eqcond_generic_indices}
\end{eqnarray}
so no impossibility can occur when computing the next coefficients 
$u_2,v_2,u_4,v_4$.
The only obstruction arises from the Fuchs index $i=6$,
which generates two necessary conditions
(one for each sign $\varepsilon$)
for the absence of movable logarithms
(again we skip the details of this classical computation),
\begin{eqnarray}
& & {\hskip -10.0 truemm}
Q_6  \equiv \cz (\Adim D - 3 d c_0) 
\left(\tildeK_1 -\frac{2 D u_0}{c v_0} \tildeK_2\right) =0,\ 
\label{Q6eps}
\\ & & {\hskip -10.0 truemm}
\tildeK_1=c K_1 - a (b -2 d) c_0 \frac{\cz^2}{c},\
\tildeK_2=  K_2 +  \left((b-2 d)^2 c_0^2 + 2 a c\right) \frac{\cz^2}{2 c^2}.
\end{eqnarray}
This defines three subcases,
\begin{eqnarray}
   & & 
\cz=0,
\label{Q6zer}
\\ & &
(b-d) c_0=0, 
\label{Q6one}
\\ & &
\tildeK_{1}= \frac{2 D u_0}{c v_0} \tildeK_2. 
\label{Q6two_one_series}
\end{eqnarray}

The first two subcases are independent of the sign $\varepsilon$,
and the third subcase can be enforced
either for one sign (condition (\ref{Q6two_one_series})) or for both signs,
leading to the stronger condition
\begin{eqnarray}
   & & 
\tildeK_{1}=0,\ \tildeK_{2}=0.
\label{Q6two_two_series}
\end{eqnarray}

For the first two cases, a first integral exists,
\begin{eqnarray}
(b-d) c_0 \cz =0:\   
K_6 &=&
\mu \cnl \left(c^2 {u'}^2 + 2 c d c_0 u' v' + (a c -(b-d) d c_0^2) {v'}^2\right)
\nonumber \\ & & 
+ \cnl^2 \left(c u v^2 + \frac{(d-b) c_0}{3} v^3\right)
+ \cnl c \cz u^2
\nonumber \\ & & 
+ 2 c K_2 u + \left(2 c K_1 - 2 K_2 (b-d) c_0\right) v,
\label{eqK6}
\end{eqnarray}
and the first integral for the third subcase 
$\tildeK_{1}=\tildeK_{2}=0$, yet to be found,
is not quartic in $(u',v')$.

\section{Method to find all the elliptic solutions}
\label{sectionEllipticMethod}

For full details on the method, we refer to \cite{CM2009,CMBook}.

The input is an $N$-th order ($N\ge 2$) any degree autonomous algebraic
ordinary differential equation (ODE) admitting a Laurent series.

The output is made of all its elliptic or degenerate elliptic solutions
in closed form.

Let us first recall a classical definition.
The \textit{elliptic order} of a nondegenerate elliptic (genus one) 
function 
\cite[Chap.~18]{AbramowitzStegun}
is the number of poles,
counting multiplicity of course,
inside a period parallelogram.
It is equal to the number of zeros.
This equality breaks down under degeneracy to genus zero, 
e.g.~for the rational function $u=(\xi-a)(\xi-b)/(\xi-c)$.

The successive steps of the algorithm are \cite{CM2009}:
\begin{enumerate}
\item
Find the analytic structure of singularities
(in our case two families of movable double poles for both $u$ and $v$,
see (\ref{eqLeadingGeneric}),
one movable simple zero and, if $K_{11}\not=0$, one movable double zero for $v$, 
at least two movable simple zeros for $u$).
Deduce the total number of poles (or, if greater, of zeros) 
of the unknown function and its derivative,
here $m=4,n=6$ for $(v,v')$, more for $(u,u')$. 

\item
Compute slightly more than $(m+1)^2$ terms in each Laurent series.

\item
Choose one of the dependent variables $(u,v)$ (call it $U$) and
define the first order $m$-th degree
subequation $F(U,U')=0$ 
(it contains at most $(m+1)^2$ coefficients $a_{j,k}$),
\begin{eqnarray}
& &
F(U,U') \equiv
 \sum_{k=0}^{m} \sum_{j=0}^{2m-2k} a_{j,k} U^j {U'}^k=0,\ a_{0,m}\not=0.
\label{eqsubeqF}
\end{eqnarray}

\item
Require at least one Laurent series of $U$
to obey $F(U,U')=0$,
\begin{eqnarray}
& & {\hskip -10.0 truemm}
F \equiv \chi^{m(p-1)} \left(\sum_{j=0}^{\jmax} F_j \chi^j
 + {\mathcal O}(\chi^{\jmax+1})
\right),\
\forall j\ : \ F_j=0,
\label{eqLinearSystemFj}
\end{eqnarray}
and solve this \textbf{linear overdetermined} system for $a_{j,k}$.

\item
Integrate each resulting first order ODE $F(U,U')=0$.
\end{enumerate}

The key advantage of this method is that 
the system of equations $F_j=0$ for the unknown coefficients $a_{j,k}$
is \textit{linear} and infinitely overdetermined,
therefore quite easy to solve.
 
\section{Elliptic and degenerate elliptic solutions, generic case}
\label{sectionSolutions}
  
By generic, we mean that the fixed constants 
$a, c, b c_0, d c_0, \cz, K_1, K_2$
obey the nonvanishing conditions 
$a c \delta D \not=0$, (\ref{eqcond_generic_indices})
and only one of the three vanishing conditions 
(\ref{Q6zer}), (\ref{Q6one}), (\ref{Q6two_one_series}).

When the algorithm of section \ref{sectionEllipticMethod}
is applied to a system of ODEs such as (\ref{eqBLS28_xi}),
a key practical ingredient is to select a ``good'' variable $U$,
i.e.~one whose total number of poles (or, if greater, of zeros) 
of $U$ and $U'$ is the smallest possible.
Since this number is always smaller for $(v,v')$ ($(4,6)$) than 
for $(u,u')$,
the natural choice is $U=v$.

Moreover, the already mentioned invariance under $\xi \to -\xi$
forbids the occurrence of odd powers of $U'=v'$ 
in the subequation (\ref{eqsubeqF}).

Since $U=v$ admits two distinct Laurent series,
the search for elliptic or degenerate elliptic 
solutions splits into (step 4):
either require one Laurent series to obey 
(the odd-parity terms have been removed)
\begin{eqnarray}
& & {\hskip -10.0 truemm}
F \equiv
              {U'}^2
+ a_{3,0} U^3
+ a_{2,0} U^2
+ a_{1,0} U
+ a_{0,0}
=0, 
\label{eqsubeqOrders23}
\end{eqnarray}
or require both Laurent series to obey 
\begin{eqnarray}
F & \equiv & 
              {U'}^4 
+ a_{3,2} U^3 {U'}^2 
+ a_{6,0} U^6        
+ a_{2,2} U^2 {U'}^2 
+ a_{5,0} U^5        
+ a_{1,2} U   {U'}^2 
+ a_{4,0} U^4        
\nonumber \\ & & 
+ a_{0,2}     {U'}^2 
+ a_{3,0} U^3        
+ a_{2,0} U^2        
+ a_{1,0} U          
+ a_{0,0}            
=0.
\label{eqsubeqOrders46}
\end{eqnarray}

\subsection{Elliptic and degenerate elliptic solutions, one series}
\label{sectionSolutionsOneSeries}
  
One of the three necessary conditions 
(\ref{Q6zer}), (\ref{Q6one}), (\ref{Q6two_one_series}) must hold true.
In this section we simply denote $(u_0,v_0)$ anyone of the two values 
$(\uzero,\vzero)$ (Eqs.~(\ref{eqLeadingGenericSol}) or (\ref{equ0v0})).

In step (2) of section \ref{sectionEllipticMethod},
it is quicker to compute simultaneously both Laurent coefficients $(u_j,v_j)$
from system (\ref{eqBLS28_xiGeneric}).
Going to $j=8$ is sufficient to obtain all the coefficients in
(\ref{eqsubeqOrders23}) 
and ensure that system (\ref{eqBLS28_xiGeneric}) is indeed
a differential consequence of (\ref{eqsubeqOrders23}).

In step (4),
with the definition (\ref{eqdefvs}),
the resulting subequation (\ref{eqsubeqOrders23}) is
\begin{eqnarray}
& &
{\vs'}^2 -\frac{4}{v_0} \vs^3 + b_2 \vs^2 +b_1 \vs +b_0=0,
\label{eqsubeqOrders23v}
\\ & &        
b_2=\left(D-4 \frac{3 d c_0 - \Adim D}{\Adim-\varepsilon}\right)
    \frac{\cz}{c \mu D},
\\ 
\cz (b-d) c_0=0:\
& & 
b_1=0,\ b_0=0,\
\tildeK_{1}= 2 \frac{D u_0}{c v_0} \tildeK_2,\ 
\\ 
\cz (b-d) c_0\not=0:\
& &
b_1=\frac{1}{\cnl \mu}
 \left(\frac{8 \tildeK_2}{D (\Adim-\varepsilon)}
       -4 (3 d c_0 - \Adim D) \frac{\cz^2}{c^2}\right),\
b_0=\hbox{arb},
\nonumber
\end{eqnarray}
i.e.~one additional constraint is found in the case $\cz (b-d) c_0=0$.

Step (5) is immediate.
Indeed, subequation (\ref{eqsubeqOrders23v})
is nothing else, up to an affine transformation,
than the canonical equation of Weierstrass \cite[Chap.~18]{AbramowitzStegun},
\begin{eqnarray}
& & {\hskip -8.0 truemm}
\vs=v_0 \wp(\xi-\xi_0,g_2,g_3)+v_2- \frac{b c_0}{\cnl c} \cz, 
\label{eqv_to_wp}
\\ & & {\hskip -8.0 truemm}
{\wp'}^2= 4 \wp^3 - g_2 \wp - g_3
        =4(\wp-e_1)(\wp-e_2)(\wp-e_3),\      
\wp'' = 6 \wp^2 - \frac{g_2}{2}.
\label{eqwp}
\end{eqnarray}

The result is
\begin{eqnarray}
& & {\hskip -6.0 truemm}
\left\lbrace
\begin{array}{ll}
\displaystyle{
v=v_0 \wp - \frac{D u_0 \cz}{c^2 \cnl v_0},\
g_3=28 \frac{v_6}{v_0},
}\\ \displaystyle{
u=\frac{u_0}{v_0} 
\left(\vs - (8 d c_0 -D (3 \Adim - \varepsilon)) \frac{\cz}{2 c \cnl}\right)
}\\ \displaystyle{\phantom{1234}
 +\left(\tildeK_1 -\frac{2 D u_0}{c v_0} \tildeK_2\right) 
  \frac{}{c \cnl D \mu (\Adim - 2 \varepsilon)} \frac{v_0}{\vs},\
}\\ \displaystyle{
}\\ \displaystyle{
\cz (b-d) c_0=0:\
 \tildeK_{1}= ((3 \varepsilon-\Adim) D -4 d c_0) \frac{\tildeK_2}{2},\        
g_2=\frac{\cz^2}{12 c^2 \mu^2},\
g_3=\frac{\cz^3}{(6 c \mu)^3},\ 
}\\ \displaystyle{
\cz (b-d) c_0\not=0:\
\tildeK_{1}= \frac{2 D u_0}{c v_0} \tildeK_2,\ 
g_3=\hbox{arbitrary},\ 
}\\ \displaystyle{
\phantom{12345678901234}
g_2=\frac{1}{(\Adim-\varepsilon)^2 D^2 \mu^2}
\left(
 -\frac{8 \tildeK_2}{3}
 +(12 d c_0 -(3 \Adim+\varepsilon) D) \frac{\cz^2}{12 c^2}
 \right).
}
\end{array}
\right.
\label{eq1fresult}
\end{eqnarray}
The effective expression of $\wp$ depends on the values of $(g_2,g_3)$ in (\ref{eqwp}),
according to the identities
\cite[\S 18.12.3]{AbramowitzStegun},
\begin{eqnarray}
& & {\hskip -6.0 truemm}
\wp(z,g_2,g_3)=
\left\lbrace
\begin{array}{ll}
\displaystyle{
\hbox{doubly periodic (``cnoidal wave'') }, g_2^3-27 g_3^2 \not=0,
}\\ \displaystyle{
3 q \coth^2 (\sqrt{3 q}z) -4 q,\
g_2=12 q^2, g_3=-8 q^3,
}\\ \displaystyle{
\frac{1}{z^2},\ g_2=g_3=0.
}
\end{array}
\right.
\end{eqnarray}

For each of the three subcases 
(\ref{Q6zer}), (\ref{Q6one}), (\ref{Q6two_one_series}),
one thus obtains two closed form solutions 
(one for each sign $\varepsilon$), namely: \hfill\break
two rational solutions for the condition (\ref{Q6zer}),
\begin{eqnarray}
(\ref{Q6zer}): & & 
\vs=v=v_0 (\xi-\xi_0)^{-2},\
u=\frac{u_0}{(\xi-\xi_0)^2} - \frac{\tildeK_2}{\mu c \cnl} (\xi-\xi_0)^{2},\ 
\nonumber \\ & &
\tildeK_{1}= ((3 \varepsilon-\Adim) D -4 d c_0) \frac{\tildeK_2}{2},\
v_6=0,\
K_6=0,
\label{eqsol1f_rational}
\end{eqnarray}
two solutions rational in one exponential for the condition (\ref{Q6one}),
\begin{eqnarray}
(\ref{Q6one}): & & 
v=v_0 \left(\tau^2 - \frac{k^2}{3}\right) - \frac{D u_0 \cz}{c^2 \cnl v_0},\ 
\tau=\frac{k}{2} \tanh \frac{k (\xi-\xi_0)}{2},\
\nonumber \\ & & 
u=\frac{u_0}{v_0} 
\left(\vs + (\Adim - 3 \varepsilon)) \frac{D \cz}{6 c \cnl}\right)
- \frac{\tildeK_2}{\mu c \cnl} \frac{v_0}{\vs},\
\nonumber \\ & &
k^2=-\frac{\cz}{12 c \mu},\
\tildeK_1=(9 \varepsilon - 7 \Adim) \frac{D \tildeK_2}{6},\
v_6=\frac{v_0}{28} \frac{\cz^3}{(6 c \mu)^3},\ 
\label{eqsol1f_trigonometric}
\end{eqnarray}
and two doubly periodic solutions for the condition (\ref{Q6two_one_series}),
\begin{eqnarray}
(\ref{Q6two_one_series}): & &
v=v_0 \wp + \frac{\cz}{4 c \cnl} (12 d c_0 -D (5\Adim - \varepsilon)),\
\nonumber \\ & &
u=\frac{u_0}{v_0}
\left(\vs - (8 d c_0 -D (3 \Adim - \varepsilon)) \frac{\cz}{2 c \cnl}\right),\ 
\nonumber \\ & &
\nonumber \\ & &
g_2=\frac{-32 c^2 \tildeK_2 + \cz^2(12 d c_0 - D(3 \Adim+\varepsilon))^2}
{12 (\Adim-\varepsilon)^2 \mu^2 D^2 c^2},\
\nonumber \\ & &
g_3=\frac{28 v_6}{v_0},\
\tildeK_1=\frac{2 D u_0}{c v_0} \tildeK_2,\ 
v_6=\hbox{arbitrary}.
\label{eqsol1f_elliptic}
\end{eqnarray}

When only one Laurent series for $v$ is enforced,
the method therefore yields two solutions of each possible kind
(elliptic, rational in one exponential, rational).

\textit{Remarks}.
\begin{enumerate}
\item
If one is interested in finding only the nondegenerate elliptic
(genus one) solutions,
such as (\ref{eqsol1f_elliptic}), 
a quicker method to find them all is to combine
the present method with the conditions on the residues
as explained in \cite{Hone2005}.
\item
For the rational and trigonometric solutions,
the ODE obeyed by $u$ has the type (\ref{eqsubeqOrders46}) (degree four),
e.g.~for the rational solution $u=\alpha (\xi-\xi_0)^{-2} + \beta (\xi-\xi_0)^{2}$,
\begin{eqnarray}
& & 
\left(\alpha {u'}^2 -2 u^3 + 8 \alpha \beta u\right)^2 
- 4 \left(u^2 -4 \alpha \beta \right)^3=0.
\end{eqnarray}
This is why $u$ should not be chosen to apply the algorithm. 

\end{enumerate}
\subsection{Elliptic and degenerate elliptic solutions, two series}
\label{sectionSolutionsTwoSeries}

One of the three necessary conditions 
(\ref{Q6zer}), (\ref{Q6one}), (\ref{Q6two_two_series}) must hold true.

The subequation (\ref{eqsubeqOrders46}) is assumed to be nonfactorizable
(nonzero value for the discriminant in ${U'}^2$).
In order to determine all the $a_{j,k}$ in subequation (\ref{eqsubeqOrders46})
and ensure that system (\ref{eqBLS28_xiGeneric}) is indeed
a differential consequence of (\ref{eqsubeqOrders46}),
it is necessary and sufficient to compute the series up to $j=14$ included,
i.e.~8 terms in the series for $u$ and 8 terms in the series for $v$
if one uses the system (\ref{eqBLS28_xiGeneric}) to perform the computation.

For the condition (\ref{Q6two_two_series}),
the found subequation (\ref{eqsubeqOrders46}) is the product of
two factors like (\ref{eqsubeqOrders23}),
the solution would only represent (\ref{eqsol1f_elliptic}),
so we discard it.

For either condition (\ref{Q6zer}), (\ref{Q6one}),
one finds the unique subequation
\begin{eqnarray}
& & {\hskip -8.0 truemm}
\cz (d-b) c_0=0:\
F \equiv 
\left\lbrack {\vs'}^2 + \Gamma (\vs-\Alpha)(\vs^2-\Beta) \right\rbrack^2 - \Delta (\vs^2-\Beta)^3=0,
\label{eqsubeqall}
\end{eqnarray}
in which
the constants $(\Alpha, \Beta, \Gamma, \Delta)$ take the values
\begin{eqnarray}
& & 
\Alpha=- \frac{3 (\Adim^2-1) D}{4 c \cnl \Adim} \cz,\
\Beta=-6 \frac{\Adim^2-1}{\cnl^2(\Adim^2-2)} \tildeK_2,\ 
\Gamma=      \frac{4 \cnl \Adim}{3 \mu D (\Adim^2-1)},\ 
\\
& & 
\Delta=\left(\frac{4 \cnl}      {3 \mu D (\Adim^2-1)}\right)^2,\
\tildeK_1=\left(-2 d c_0 + \frac{D \Adim (\Adim^2-3)}{2(\Adim^2-2)} \right)\tildeK_2.
\end{eqnarray}

Because of the parity invariance, 
the general solution of (\ref{eqsubeqall}) cannot involve $\wp'$ 
and is the quotient of two second degree polynomials of $\wp$.
It can be obtained by brute force 
with the Maple command
\verb+with(algcurves);Weierstrassform(...)+ \cite{MapleAlgcurves},
but the structure of singularities allows a straightforward integration. 
Indeed, the \textit{a priori} solution
\begin{eqnarray}
& & 
\vs=\frac{a_2 \wp^2 + a_1 \wp + a_0}{(\wp -e_1)(\wp-e_2)},\ 
\end{eqnarray}
must have two double poles with principal parts 
$\vzero (\xi-\xi_\varepsilon)^{-2}$,
therefore the poles $e_1,e_2$ must be \cite[\S 18.3.1]{AbramowitzStegun}
two of the three zeros $e_j$ of $\wp'$, Eq.~(\ref{eqwp}),
\begin{eqnarray}
& & {\hskip -6.0 truemm}
\wp(\xi_j)=e_j,\ j=1,2;\
\xi \to \xi_j:\ 
\wp(\xi) -e_j 
 \sim \frac{1}{2} \left(6 e_j^2 - \frac{g_2}{2} \right) (\xi-\xi_j)^2.
\end{eqnarray}
The result,
\begin{eqnarray}
& & {\hskip -6.0 truemm}
\vs=a_2 
+ \frac{v_{0,1}}{2} \left(6 e_1^2 - \frac{g_2}{2} \right) \frac{1}{\wp-e_1}
+ \frac{v_{0,2}}{2} \left(6 e_2^2 - \frac{g_2}{2} \right) \frac{1}{\wp-e_2},
\end{eqnarray}
in which $v_{0,1}, v_{0,2}$ are the two values of $\vzero$, 
Eq.~(\ref{eqLeadingGenericSol}),
can be written as
\begin{eqnarray}
& & 
\frac{3 \mu D}{\cnl}\vs=
3 \Adim e_3 + e_1-e_2
- (e_1-e_2) \frac{(2 \wp +e_3+\Adim (e_1-e_2))(\wp-e_3)}{(\wp-e_1)(\wp-e_2)},\
\nonumber \\ & &
e_1+e_2=-e_3=\frac{\cz}{12 \mu c},\
(e_1-e_2)^2=\frac{2}{3 \Adim^2 \mu^2 (\Adim^2-2)} \tildeK_2 
            +\left(\frac{\cz}{4 \mu c}\right)^2,\
\nonumber \\ & &            
\\ & &
g_2=\frac{\cz}{6 \mu c} \frac{\tildeK_2}{3 \Adim^2 \mu^2 (\Adim^2-2)}
    +   \left(\frac{\cz}{6 \mu c}\right)^3,\
g_3=                  \frac{2 \tildeK_2}{3 \Adim^2 \mu^2 (\Adim^2-2)}
    + 3 \left(\frac{\cz}{6 \mu c}\right)^2.
\nonumber 
\end{eqnarray}
This represents two solutions
since $e_1-e_2$ can take two opposite values.

This solution is truely elliptic,
except when two of the three roots $e_1,e_2,e_3$ are equal.
Finally, the value of $u$ is easily obtained from the second equation
of system (\ref{eqBLS28_xiGeneric}).

\section{Conclusion and perspectives}
  
When the coefficients of the Boussinesq system
(\ref{eqBLS28}) take generic values 
as defined at the beginning of section \ref{sectionSolutions},
we have obtained in closed form 
\textit{all} the singlevalued solutions which are 
either elliptic (``cnoidal'')
or rational in one exponential
(this includes polynomials of $\tanh$ and $\sech$)
or rational.
Naturally,
one then must select those solutions which make the internal
waves real and bounded.
 
When the quadratic nonlinearities in the Boussinesq system
(\ref{eqBLS28}) are insufficient
and cubic nonlinearities must be incorporated,
the resulting system
\cite[Eq.~(50)]{NguyenDias2008}
can be handled similarly.
The various Boussinesq systems displayed in 
\cite{BLS2008} could similarly be processed.
 
\section*{Acknowledgements}

RC thanks the WASCOM organizers for invitation.
Partial financial support has been provided by the Hong Kong
Research Grants Council contract
HKU 7038/07P. 


\vfill \eject

\begin{thebibliography}{99}

\bibitem{AbramowitzStegun} M.~Abramowitz, I.~Stegun,
\textit{Handbook of mathematical functions},
Tenth printing (Dover, New York, 1972).

\bibitem{BLS2008} J.L.~Bona, D.~Lannes and J.-C.~Saut,      
Asymptotic models for internal waves,
J.~Math.~Pures Appl.~{\bf 89} (2008) 538--566.

\bibitem{BriotBouquet} C.~Briot et J.-C.~Bouquet,       
{\it Th\'eorie des fonctions elliptiques},
1\`ere \'edition  (Mallet-Bachelier, Paris, 1859);
2i\`eme \'edition (Gauthier-Villars, Paris, 1875).
\verb+http://gallica.bnf.fr/document?O=N099571+ 

\bibitem{Chen1998} M.~Chen,                               
Exact solutions of various Boussinesq systems,
Appl.~Math.~Lett.~{\bf 11} (1998) 45--49.

\bibitem{CMBook} R.~Conte and M.~Musette,
\textit{The Painlev\'e handbook} (Springer, Berlin, 2008).

\bibitem{CM2009} R.~Conte and M.~Musette,
Elliptic general analytic solutions,
Studies in Applied Mathematics {\bf 123} (2009) 63--81.
http://arxiv.org/abs/0903.2009

\bibitem{MapleAlgcurves} Mark van Hoeij,
package ``algcurves'', Maple V (1997).
\hfill\break\noindent
 http://www.math.fsu.edu/\~{ }hoeij/algcurves.html

\bibitem{Hone2005} A.N.W.~Hone,                                  
Non-existence of elliptic travelling wave solutions of the complex
Ginzburg-Landau equation,
Physica D {\bf 205} (2005) 292--306.

\bibitem{MC2003} M.~Musette and R.~Conte,
Analytic solitary waves of nonintegrable equations,
Physica D {\bf 181} (2003) 70--79.
http://arXiv.org/abs/nlin.PS/0302051

\bibitem{NguyenDias2008} Nguyen Hai Yen and Fr\'ed\'eric Dias,
A Boussinesq system for two-way propagation of interfacial waves,
Physica D {\bf 237} (2008) 2365--2389.


\end{thebibliography}
\end{document}